\documentclass[aps,twocolumn,prc,floatfix,showpacs,preprintnumbers,amsmath,amssymb,nofootinbib,groupedaddress]{revtex4-1}
\usepackage[normalem]{ulem}
\usepackage{wasysym}
\usepackage{color}
\usepackage{graphicx}
\usepackage{dcolumn}    
\usepackage{bigstrut}
\usepackage{amsmath,amsfonts,amsthm,bm}
\usepackage{CJK}
\usepackage[pdfstartview=FitH,
            CJKbookmarks=true,
            bookmarksnumbered=true,
            bookmarksopen=true,
            colorlinks,
            pdfborder=001,
            linkcolor=blue,
            anchorcolor=blue,
            citecolor=blue
            ]{hyperref}

\begin{document}
\begin{CJK}{UTF8}{gbsn}
%
\title{Harmonic potential theorem: \\extension to spin-, velocity- and density-dependent interactions}

\author{S. Zanoli}
\affiliation{Dipartimento di Fisica ``Aldo Pontremoli'', Universit\`a degli Studi di Milano, 20133 Milano, Italy}
\affiliation{INFN,  Sezione di Milano, 20133 Milano, Italy}

\author{X. Roca-Maza}
\email{xavier.roca.maza@mi.infn.it}
\affiliation{Dipartimento di Fisica ``Aldo Pontremoli'', Universit\`a degli Studi di Milano, 20133 Milano, Italy}
\affiliation{INFN,  Sezione di Milano, 20133 Milano, Italy}

\author{G. Col\`o}
\affiliation{Dipartimento di Fisica ``Aldo Pontremoli'', Universit\`a degli Studi di Milano, 20133 Milano, Italy}
\affiliation{INFN,  Sezione di Milano, 20133 Milano, Italy}

\author{Shihang Shen (申时行)}
\affiliation{Dipartimento di Fisica ``Aldo Pontremoli'', Universit\`a degli Studi di Milano, 20133 Milano, Italy}
\affiliation{INFN,  Sezione di Milano, 20133 Milano, Italy}

\date{\today} 

\begin{abstract}
  One of the few exact results for the description of the time-evolution of an inhomogeneous, interacting many-particle system is given by the Harmonic Potential Theorem (HPT) \cite{dobson1994}. The relevance of this theorem is that it sets a tight constraint on time-dependent many-body approximations. In this contribution, we show that the original formulation of the HPT is valid also for the case of spin-, velocity- and density-dependent interactions. This result is completely general and relevant, among the rest, for nuclear structure theory both in the case of {\em ab initio} and of more phenomenological approaches.
As an example, we report on a numerical implementation by testing the small-amplitude limit of the time-dependent Hartree-Fock -- also known as Random Phase Approximation (RPA) -- for the translational frequencies of a neutron system trapped in a harmonic potential.
\end{abstract}

\pacs{21.60.Jz, 71.45.Gm, 73.20.Mf, 21.10.Re  
} 

\maketitle 

The Harmonic Potential Theorem (HPT) \cite{dobson1994} -- an extension of the Kohn's theorem \cite{kohn1961} and further generalizations \cite{brey1989,yip1991} -- is one of the few exact results for the description of the time-evolution of an inhomogeneous, interacting many-particle system. Specifically, it describes the motion of such a system, when confined in a parabolic potential well, under the action of a spatially uniform time-dependent external field. The system displays a harmonic motion of all particles oscillating as a whole. Its frequency coincides with the trapping harmonic oscillator frequency, regardless of the interparticle interaction. This result is based on the invariance of the harmonic potential under a transformation to a homogeneously  accelerated  reference frame \cite{vignale1995}: {\it the center of mass is completely decoupled from the internal degrees of freedom}. 

It must be stressed that this sets an interesting constraint on approximate time-dependent many-body theories \cite{onida2002}. In particular, the time-dependent local density approximation (TDLDA) of the time-dependent density functional theory satisfies the HPT \cite{dobson1994}. This is essentially because the exchange-correlation potential is local in time and space. The Gross and Kohn approximation \cite{gross1985} violates the HPT instead, but it has been shown that by introducing some modifications on the exchange-correlation potential it can satisfy this theorem 
\cite{vignale1996}.  

In its original formulation, the HPT assumes a two-body force that depends only on the relative coordinates of the interacting particles. However, in different physical systems, the spin-, velocity- or density-dependence of the interaction can be crucial for a realistic description of the observed phenomenology. For example, the nucleon-nucleon interaction is strongly spin-dependent, and produces a bound state for the neutron-proton system (the deuteron) with aligned spins ($S$ = 1), while  all two-nucleon $S$ = 0 configurations are known to be unbound. Spin-dependent interactions are also important to describe magnetic phenomena \cite{katsnelson2008,jungwirth2014}. 

If we deal with systems characterized by short-range interactions (see for example \cite{flenner2014,frerot2018} among others), the associated non-local equations (e.g., the Hartree-Fock equations) may become quite complicated. If the interaction can be turned into a contact one, these non-local equations can become local, and this represents a practical, and often quite accurate, alternative. With this aim, the so-called density matrix expansion was developed in Refs. \cite{negele1972,negele1975}. Such a method is based on the expansion of the non-local one-body density $\rho({\bm r},{\bm r}^\prime)$ of the system under study, around $({\bm r}+{\bm r}^\prime)/2$, up to the needed order in powers of ${\bm r}-{\bm r}^\prime$. 
This brings in derivatives acting on the wave functions evaluated in $({\bm r}+{\bm r}^\prime)/2$;
in other terms, when looking at the direct and exchange matrix elements of the original 
interaction, one can realize that these have been mapped onto those of a contact, velocity-dependent interaction (see e.g. Sec. I.D of Ref. \cite{skyrme1}).
 
Finally, three-body interactions have been shown to be of paramount importance in different fields of  phyiscs \cite{hammer2013,horinouchi2015,ringschuck}. 
In the {\em ab initio} approaches to nuclear structure, three-body forces and possibly induced four-body or higher-body forces do show up 
\cite{RevModPhys.81.1773}.
Three-body interactions are very complicated to deal with: they require an extension of the usual quantum many-body techniques. One possibility to address this issue, followed within current nuclear density functional approaches, is to modelize a three-body interaction by adopting an effective two-body density-dependent interaction \cite{skyrme1958}. In this way, 
the three-body force can be seen as having been averaged on the density of one of the particles.
Therefore, in the light of this discussion, one may deem necessary to generalize the HPT, in order to set exact constraints on time-dependent many-body theories based on spin-dependent, velocity-dependent and three-body/higher-body forces (or density-dependent two-body forces).     

The HPT was established by John F. Dobson in Ref. \cite{dobson1994}. It starts from a general $N$-particle Hamiltonian under the action of an external time-dependent and homogeneous field of the type $-{\bm F}(t)$,
\begin{equation}
\mathcal{H}(\{{\bm r}_i\}) = \mathcal{H}_0 - {\bm F}(t)\cdot\sum_{j=1}^N {\bm r}_j \ ,
\label{eq1}  
\end{equation}
where $\{{\bm r}_i\}={\bm r}_1,\dots,{\bm r}_N$ and $\mathcal{H}_0$ corresponds to the unperturbed Hamiltonian for the $N$ interacting particles that are trapped in a harmonic potential, that is,   
\begin{equation}
\mathcal{H}_0 = \sum_{i=1}^N\left[-\frac{\hbar^2{\bm \nabla_i}^2}{2m}+\frac{1}{2}{\bm r}_i\cdot {\bm K}\cdot{\bm r}_i\right] +  \frac{1}{2}\sum_{j\ne k=1}^NV(\{\vert{\bm r}_j-{\bm r}_k\vert\}) \ ,
\label{eq2}  
\end{equation}
where ${\bm K}$ is the spring-constant matrix of the harmonic trap. It is important to mention that a suitable choice of ${\bm K}$ may allow one to modelize very different physical systems such as non-neutral quantum wires or dots \cite{maksym1990,broido1990}, Hooke's atoms \cite{taut1993,filippi1994}, Hooke's species \cite{pan2003}, or spherical nuclei \cite{ringschuck}, among others. In the original formulation $V(\{\vert{\bm r}_j-{\bm r}_k\vert\})$ was an arbitrary two-particle potential which depends on the relative coordinates of particles $j$ and $k$. 

In the present contribution, we generalize Eq. (\ref{eq1}) and consider an arbitrary velocity- and density-dependent interaction that preserves, as it should, Galilean invariance. In addition, we trivially generalize $V(\{\vert{\bm r}_j-{\bm r}_k\vert\})$ to depend on spin ${\bm \sigma}$ and isospin ${\bm \tau}$ as well,
\begin{eqnarray}
&& V = V_C(\{\vert{\bm r}_j-{\bm r}_k\vert\})+V_S(\{\vert{\bm r}_j-{\bm r}_k\vert\}){\bm \sigma}_j\cdot{\bm \sigma}_k+\nonumber\\ &&+V_T(\{\vert{\bm r}_j-{\bm r}_k\vert\}){\bm \tau}_j\cdot{\bm \tau}_k + 
V_{ST}(\{\vert{\bm r}_j-{\bm r}_k\vert\}){\bm \sigma}_j\cdot{\bm \sigma}_k {\bm \tau}_j\cdot{\bm \tau}_k 
\ . \nonumber \\
\label{eq3}  
\end{eqnarray}

To study the time evolution of such a 
system, 
given the specific type of external perturbation in Eq. (1), the HPT considers a position-independent and time-dependent shift ${\bm x}(t)$ of the wave function $\Psi_0$ that is solution of the unperturbed Hamiltionian. 
In mathematical form, the time-evolved wave function can be written as follows,    
\begin{equation}
\Psi_{\rm HPT}(\{{\bm r}_i\},t) = e^{-\imath \frac{E_0}{\hbar}t - \imath NS(t) + \imath\frac{N}{\hbar}m\frac{d{\bm x}}{dt}\cdot {\bm R}}\Psi_0(\{\bar{\bm r}_i\}) \ ,
\label{eq4}  
\end{equation}
where $\bar{\bm r}_j\equiv {\bm r}_j - {\bm x}(t)$, ${\bm R}\equiv\frac{1}{N}\sum_{j=1}^N {\bm r}_j$ and the phase $S(t)$ is defined as
\begin{equation}
S(t) = \frac{1}{\hbar}\int_0^t\left[\frac{1}{2}m\dot{\bm x}(t^\prime)^2-\frac{1}{2}{\bm x}(t^\prime)\cdot{\bm K}\cdot{\bm x}(t^\prime) \right]dt^\prime \ .
\label{eq5}  
\end{equation}
We note that $\Psi_0$ is stationary when referred to the accelerated frame $\bar{\bm r}$ and that the phase shift $S(t)$ brings it back to the rest frame ${\bm r}$ \cite{dobson1994}. $E_0$ is the 
corresponding eigenenergy.

The original HPT proofs that $\Psi_{\rm HPT}(\{{\bm r}_i\},t)$, as written in Eq. (\ref{eq4}), is a solution of the time-dependent many-body Schr\"odinger equation
\begin{equation}
\mathcal{H}(\{{\bm r_i}\},t) \Psi_{\rm HPT}(\{{\bm r}_i\},t) = i\hbar\frac{\partial}{\partial t}\Psi_{\rm HPT}(\{{\bm r}_i\},t)\ ,
\label{schrodinger}  
\end{equation}  
provided that ${\bm x}(t)$ follows the classical harmonic oscillator equation (cf. Appendix B of Ref. \cite{sahni2016}),
\begin{equation}
  m\ddot{\bm x}=-{\bm K}\cdot {\bm x} + {\bm F}(t) \ .
  \label{hpt}
\end{equation}
Later, it has been shown that the HPT wave function can be derived from first principles via the Feynman Path Integral method \cite{li2013, feynman1948} and the interaction representation of quantum mechanics \cite{lai2015}. Following, for example, the proof via the {\it operator method} (Appendix B.1 of Ref. \cite{sahni2016}), one realizes that 
the interactions that depend on the spin and/or isospin do not modify the proof of the HPT. Specifically, the general interaction (\ref{eq3}) proposed here does not modify the value of the commutators shown in Eqs. (B16-B19) of Ref. \cite{sahni2016},
\begin{eqnarray}
\left[\frac{i}{\hbar}Nm\dot{\bm x}\cdot{\bm R}, \mathcal{H}_0\right] &=& - N\dot{\bm x}\cdot{\bm P},  \label{eq6} \\ 
\left[\frac{i}{\hbar}Nm\dot{\bm x}\cdot{\bm R}, \left[ \frac{i}{\hbar}Nm\dot{\bm x}\cdot{\bm R} ,\mathcal{H}_0\right]\right] &=& Nm\dot{\bm x}^2,  \label{eq7}\\ 
\left[-\frac{i}{\hbar}N{\bm x}\cdot{\bm P}, \mathcal{H}_0\right] &=& - N{\bm x}\cdot{\bm K}\cdot{\bm R}, \ \ \label{eq8} \\
\left[-\frac{i}{\hbar}N{\bm x}\cdot{\bm P}, \left[-\frac{i}{\hbar}N{\bm x}\cdot{\bm P},\mathcal{H}_0\right]\right] &=& N{\bm x}\cdot{\bm K}\cdot{\bm x}, 
\label{eq9}
\end{eqnarray}
which are at the center of the proof. Note that we have defined ${\bm P}\equiv \sum_j^N {\bm p}_j\big/N$, being ${\bm p_j}$ the conjugate variables with respect to ${\bm r_j}$.

The interaction $V$ of any non-relativistic system should preserve Galilean invariance. 
For a velocity- or momentum-dependent interaction, 
the simplest combination of the momenta that preserves Galilean invariance is ${\bm p}_i-{\bm p}_j$. Hence, it is immediate to show that commutators in Eqs. (\ref{eq6}-\ref{eq7}) will not change, as    
\begin{equation}
[\frac{i}{\hbar}Nm\dot{\bm x}\cdot{\bm R}, {\bm p}_i-{\bm p}_j] = \frac{i}{\hbar}m\sum_{k=1}^N[\dot{\bm x}\cdot{\bm r_k}, {\bm p}_i-{\bm p}_j] =0 \ . 
\label{galilean-v}  
\end{equation}
More generally, it has been verified that, in the case of any Galilean invariant interaction $V$,  
\begin{equation}
[f({\bm r}),V] = 0 \ ,
\label{eq10}  
\end{equation}
for any local operator $f({\bm r})$ that is only function of the spatial coordinates
(cf. Eq. (6.4) of Ref. \cite{lipparini1989}). We note that the latter expression is not only valid for zero-range velocity-dependent forces, but also for finite-range forces with exchange terms. Regarding commutators in Eqs. (\ref{eq8}-\ref{eq9}), they are trivially unchanged. It must be stressed that 
three- (or many-) body forces depending on the relative coordinates of the involved particles would  not modify the proof of the HPT [i.e., commutators in Eqs. (\ref{eq6}-\ref{eq9}) remain as they are].

Often, in nuclear physics, it has become customary to take into account medium effects by adopting effective two-body, density-dependent forces. As an example, within a Hartree-Fock (HF) calculation for an even-even nucleus, one can show that a zero-range three-body interaction of the type 
\begin{equation}
 V^{(3)} = g\delta({\bm r}_1-{\bm r}_2)\delta({\bm r}_2-{\bm r}_3) \ 
\label{eq11}
\end{equation}
is equivalent to a two-body density-dependent force of the form \cite{vautherin1972}
\begin{eqnarray}
  V^{(3)} &=& g\frac{1+P_\sigma}{6}\delta({\bm r}_1-{\bm r}_2)\rho\left(\frac{{\bm r}_1+{\bm r}_2}{2}\right), 
\label{eq12}
\end{eqnarray}
where $P_{\sigma}$ is the exchange operator between particles 1 and 2 in the spin space. This equivalence is not valid, strictly speaking, beyond HF \cite{bogner2010}; density-dependemt forces should be taken as a mere phenomenological way to mimic many-body effects. 
It has been shown, since a few decades, that a fractional power ($\alpha < 1$) of the density ($\rho^\alpha$) is more appropriate if one wishes to accurately describe at the same time  
nuclear bulk properties and nuclear excitations 
(cf., for instance, \cite{paar2007,bennaceur2014,roca-maza2018b}). 
$\alpha < 1$ is needed for a realistic description of the nuclear incompressibility 
\cite{PhysRevC.70.024307}.
 
For density-dependent forces such as the one in Eq.~(\ref{eq12}), one needs to evaluate the corresponding part of the commutators in Eqs. (\ref{eq6}-\ref{eq9}) assuming that $\mathcal{H}_0$ 
explicitly depends on the one-body density,
\begin{eqnarray}\label{def_dens}
\rho({\bm r},t)&\equiv&\frac{1}{N}\int d{\bm r_2}\dots d{\bm r_N} \Psi^\dag({\bm r}, {\bm r}_2\dots{\bm r}_N,t)\Psi({\bm r}, {\bm r}_2\dots{\bm r}_N,t).\nonumber\\
\end{eqnarray}
Here, $\Psi$ labels the general many-body wave function. In the case of the HPT wave function (\ref{eq4}), 
the density of the system is
invariant under the solid shift ${\bm x(t)}$, 
that is, $\rho({\bm r},t)
= \rho(\bar{\bm r})$, where $\rho(\bar{\bm r})$ is the static one-body density solution of $\mathcal{H}_0(\bar{\bm r})$. 
Hence, $[\mathcal{H}_0(\bar{\bm r}),\rho(\bar{\bm r})]=0$. Expanding the latter commutator expression one finds
\begin{widetext}  
\begin{eqnarray}
  [\mathcal{H}_0(\bar{\bm r}),\rho(\bar{\bm r})]&=&0\nonumber \\
  &=&\left[\sum_{j=1}^N\frac{\bar{\bm p}_j^2}{2m},\rho(\bar{\bm r})\right]+\frac{1}{2}\left[\sum_{j=1}^N \bar{\bm r}_j\cdot{\bm K}\cdot\bar{\bm r}_j,\rho(\bar{\bm r})\right] + 
  \frac{1}{2} \left[ \sum_{j \ne k = 1}^N 
  V(\{\vert\bar{\bm r}_j-\bar{\bm r}_k\vert\}),\rho(\bar{\bm r})\right] \label{eq13} \\
  &=& \left[\sum_{j=1}^N\frac{({\bm p}_j-{\bm p}_x)^2}{2m},\rho(\bar{\bm r})\right]=\sum_{j=1}^N\frac{1}{2m}\Big(\left[{\bm p}_j^2,\rho(\bar{\bm r})\right]+\left[{\bm p}_x^2,\rho(\bar{\bm r})\right]-2\left[{\bm p}_x\cdot{\bm p}_j,\rho(\bar{\bm r})\right]\Big)\ ,
  \label{eq14} 
\end{eqnarray}
\end{widetext}
where ${\bm p}_x=i\hbar\partial/\partial {\bm x}$. The second commutator at the r.h.s. of Eq. (\ref{eq13}) is trivially zero, and the third one is also zero due to Eq. (\ref{eq10}). The three commutators at the r.h.s. of Eq. (\ref{eq14}) can be evaluated as follows. We define a unitary transformation so that $\mathcal{T}\Psi_0(\{{\bm r}\})=\Psi_0(\{\bar{\bm r}\})$, namely $\mathcal{T}=\exp\left[-iN{\bm x}(t)\cdot{\bm P}\right]$. Therefore, the first commutator is
\begin{equation}
  \left[{\bm p}_j^2,\rho(\bar{\bm r})\right]=\left[{\bm p}_j^2,\Psi_0^\dag(\{{\bm r}\})\mathcal{T}^\dag\mathcal{T}\Psi_0(\{{\bm r}\})\right]=\left[{\bm p}_j^2,\rho({\bm r})\right]=0,  
\label{eq15}
\end{equation}
since $[\mathcal{H}_0({\bm r}),\rho({\bm r})]=0$ [we have simplified the notation, by omitting the pre-factor and the integral of Eq. (\ref{def_dens})]. The second one is
\begin{equation}
  \left[{\bm p}_x^2,\rho(\bar{\bm r})\right]=\left[{\bm p}_x^2,\Psi_0^\dag(\{{\bm r}\})\mathcal{T}^\dag\mathcal{T}\Psi_0(\{{\bm r}\})\right]=\left[{\bm p}_x^2,\rho({\bm r})\right]=0,   
\label{eq16}
\end{equation}
since ${\bm p}_x$ commutes with ${\bm r}$. The last commutator should be zero because of Eq. (\ref{eq14}):
\begin{eqnarray}
  \left[{\bm p}_x\cdot{\bm p}_j,\rho(\bar{\bm r})\right]&=&\left[{\bm p}_x\cdot{\bm p}_j,
  \Psi_0^\dag(\{{\bm r}\})\mathcal{T}^\dag\mathcal{T}\Psi_0(\{{\bm r}\})\right]\nonumber\\
  &=&\left[{\bm p}_x\cdot{\bm p}_j,\rho({\bm r})\right]=0. 
\end{eqnarray}
This implies that the commutator of ${\bm p}_j$ projected along the direction of ${\bm x}(t)$ commutes with $\rho({\bm r})$, ensuring that Eqs. (\ref{eq8}-\ref{eq9}) remain valid in the case of density-dependent forces as well. For the case of density-dependent forces the commutators in Eqs. (\ref{eq6}-\ref{eq7}) are trivially unchanged.  

\begin{table}[!th]
\caption{Total binding energy with respect to the Thomas-Fermi solution of Eq. (\ref{eq18}), root mean square radius, excitation energy of the translational mode with respect to the trap frequency, and fraction of the model-independent energy-weighted sum rule exhausted by the mode (see text), for different neutron drops ranging from 2 to 50 neutrons, trapped in an harmonic oscillator of $\hbar\omega_{\rm trap}=10$ MeV.}
\label{tab1}
\centering
\begin{tabular}{rccccc}
\hline\hline
N & $E/E_{\rm TF}$& $\langle r^2\rangle^{1/2}$& $\omega_{\rm RPA}/\omega_{\rm trap}$ & $m_1/m_1^{\rm D.C.}$\\
  &             &[fm]& &[\%]\\
\hline  
 2 & 0.844 & 2.22 & 1.000 & 99.99\\
 8 & 0.723 & 2.63 & 1.002 & 99.98\\
16 & 0.714 & 2.95 & 1.003 & 99.98\\
20 & 0.685 & 3.07 & 1.004 & 99.95\\
40 & 0.677 & 3.51 & 1.005 & 99.61\\
50 & 0.685 & 3.65 & 1.007 & 99.89\\
\hline\hline
\end{tabular}
\end{table}

Given all previous discussions, it is now evident that the small-amplitude limit of the time-dependent Hartree-Fock theory, 
commonly known as Random Phase Approximation (RPA), preserves the HPT also in the case of spin- velocity- and density-dependent forces.
In what follows, we will numerically show that a system of neutrons in an isotropic harmonic trap (with ${\bm K}_{ii}=m\omega^2$ and ${\bm K}_{ij}=0$), solved within the RPA with an effective zero-range interaction which is spin-, velocity- and density-dependent, satisfies the HPT. This system is dubbed 
{\em neutron drop} and it is a useful benchmark for testing nuclear models 
\cite{gandolfi2011, maris2013,potter2014,shen2018}. 
RPA is a very successful approach for different types of fermionic systems; in nuclear physics it 
is the tool of choice for studying the collective motion, also in connection with
the extraction of the parameters governing the nuclear equation of state, or with 
applications to processes of interest for particle physics and astrophysics \cite{ringschuck, paar2007, roca-maza2018b}. 

As our code is in spherical symmetry, we should seek among the RPA solutions 
with angular momentum and parity $J^\pi=1^-$ . The translational modes are known
\cite{ringschuck} to be excited by the so-called isoscalar dipole operator, 
$\mathcal{O}=\sum_{i=1}^N r_i Y_{10}(\hat{\bm r}_i)$. 
The mode we are after should be essentially the only one excited by this operator, and its
frequency should be equal to the trap frequency $\omega$ according to the HPT. To better
characterize its translational nature, we can look at its transition density. 
Transition densities are defined, for any given RPA state $n$, by
\cite{ringschuck,colo2013}
\begin{equation}
\delta\rho({\bm r},t)=\frac{1}{N}\int d{\bm r_2}\dots d{\bm r_N} \Psi^\dag_{n}({\bm r}, {\bm r}_2\dots{\bm r}_N,t)\Psi_0({\bm r}, {\bm r}_2\dots{\bm r}_N).
\end{equation}
If the motion is associated with an infinitesimal displacement equal to ${\bm A}$, the transition density reads
\cite{dobson1994}
\begin{equation}
\delta \rho({\bm r}) = -{\bm A}\cdot{\bm \nabla}\rho_0({\bm r}).
\label{eq17}  
\end{equation}
Here, $\rho_0$ is the ground state density and the displacement ${\bm A}$ must be related
to the harmonic motion in the trap ${\bm x}(t)$. If
${\bm x}(t)={\bm A}cos(\omega t+\phi)$, then $A=\sqrt{\frac{2}{Nmc^2 \hbar\omega}}\hbar c$ because the energy $\hbar\omega$ 
should be equal to the classical energy.

For our calculations we have used the RPA code published in Ref. \cite{colo2013}, adapted to treat a system in a harmonic trap. This code has been implemented with a two-body interaction of the Skyrme type which is zero-range, spin-, velocity- and density-dependent in its standard form \cite{skyrme1}. For the numerical implementation, we have picked up the SAMi parameterization of the Skyrme model \cite{roca-maza12b}. 

\begin{figure}[t!]
\vspace{5mm}  
\includegraphics[width=\linewidth,clip=true]{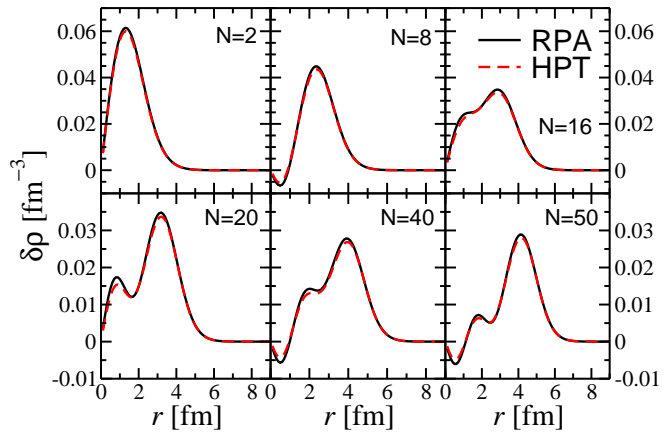}
\caption{Transition densities obtained from the RPA calculation (full line) and from the HPT in the 
form of Eq. (\ref{eq20}) (dashed line), in the case of different neutron drops trapped in a harmonic potential with $\hbar\omega=10$ MeV. 
}
\label{fig1}
\end{figure}

In Table \ref{tab1}, we show some results for different neutron drops ranging from 2 to 50 neutrons trapped in a harmonic oscillator potential having $\hbar\omega_{\rm trap}=10$ MeV. In the second column, the total energy with respect to the Thomas-Fermi solution for a non-interacting $N$-fermion system \cite{ringschuck},
\begin{equation}
E_{\rm TF} = \frac{3^{4/3}}{4}\hbar\omega_{\rm trap} N^{4/3} \ ,   
\label{eq18}
\end{equation}
is given. In the third column, the predictions for the root mean square neutron radius are shown. The results for the total binding energy and radius are consistent with previous calculations available in the literature \cite{gandolfi2011, maris2013,potter2014,shen2018}, and are the only ones shown here that depend on the interaction. They are provided for the sake of completeness. In the fourth column, the frequency of the translational mode 
as found in the RPA calculations is given with respect to the trap frequency. Finally, in the last column, the fraction of the model independent energy-weighted sum rule (EWSR) exhausted by the mode is shown \cite{ringschuck}. The EWSR can be analytically calculated from the double-commutator (D.C.) as
\begin{eqnarray}
  m_1^{\rm D.C.} &=& \frac{1}{2}\langle\Psi_0^{\rm HF}\vert[\mathcal{O},[\mathcal{H}_0,\mathcal{O}]]\vert\Psi_0^{\rm HF}\rangle\nonumber\\
&=& -\frac{1}{2}\langle\Psi_0^{\rm HF}\vert[\mathcal{O},[\frac{\hbar^2{\bm \nabla}^2}{2m},\mathcal{O}]]\vert\Psi_0^{\rm HF}\rangle\nonumber\\
  &=&\frac{9\hbar^2}{8\pi m}N,
\label{eq19}
\end{eqnarray}
in a model independent fashion. In fact, due to Eq. (\ref{eq10}), the kinetic energy is the only term contributing to $m_1^{\rm D.C.}$. It is clear from the table that the sharp RPA mode coincides, within $\permil$ accuracy, with the trap frequency and that such mode is the only one appreciably excited in the RPA: it exhausts essentially all the $m_1^{\rm D.C.}$. This is a powerful test for the extended HPT that has been discussed in this work.

In Fig. \ref{fig1}, we compare the RPA transition densities $\delta\rho_{\rm RPA}$ 
with the expected result from the HPT given in  (\ref{eq17}): this becomes, in spherical symmetry,
\begin{equation}
\delta \rho_{\rm HPT}(r) = -\sqrt{\frac{2}{Nmc^2 \hbar\omega}}\hbar c \frac{d\rho_0}{dr} \ .
\label{eq20}
\end{equation}
The results show a very good numerical agreement 
between the calculation and the expectations from the HPT.
The root mean square deviation between the two results shown in Fig. \ref{fig1} is around, or smaller than, $6\times 10^{-4}$ fm${}^{-3}$ 
which corresponds to a numerical error at the 1\% level or below. 
This confirms 
that the RPA approach 
based on spin-, velocity- and density-dependent Hamiltonian satisfies 
the extended HPT theorem that we have demonstrated here.

In summary, we have extended the HPT to spin-, velocity- and density-dependent interactions. 
This generalization is of fundamental relevance. 
We had chiefly in mind the case of the atomic nucleus, and we have used a system of
neutrons to demonstrate that the generalized HPT can be fulfilled numerically with high
accuracy. This was done in the case of a specific Hamiltonian. Nevertheless, in keeping with the steady progress of
{\em ab initio} approaches to nuclear structure \cite{Gandolfi2016,PhysRevLett.122.042501} one can be
confident that the generalized HPT may be relevant for this domain (cf. also \cite{PhysRevLett.118.252501} and references therein). In addition,
there exist other types of physical systems that 
are governed by spin-, velocity-, or density-dependent interactions. If they are
composed by many fermions, such systems are difficult to be fully understood from a microscopic point of view. Hence, this extension of the theorem enables to test approximate time-dependent many-body theories dealing with the description of the time-evolution of an inhomogeneous, interacting many-particle system, by setting a firm constraint.

The authors are greatful to G. Onida for useful comments and a careful reading of the manuscript. Funding from the European Union's Horizon 2020 research and innovation programme under grant agreement No 654002 is acknowledged.

\end{CJK}

\bibliography{bibliography.bib}

\end{document}